\documentclass[%
reprint,
amsmath,amssymb,
aps,
pra,
floatfix,
]{revtex4-2}
\usepackage{amsmath}
\usepackage{amssymb}
\usepackage{graphicx}
\usepackage{epsfig}
\usepackage{epstopdf}
\usepackage{dcolumn}
\usepackage{bm}
\usepackage{array}
\usepackage{mathtools}
\usepackage{lipsum}
\usepackage{booktabs}
\usepackage{siunitx}
\usepackage{enumitem}
\usepackage{url}
\usepackage{physics}
\usepackage[singlelinecheck=false,justification=justified,font=small]{subcaption}   
\usepackage[colorlinks,bookmarks=false,citecolor=blue,linkcolor=blue,urlcolor=blue]{hyperref}

\begin{document}

\title{Electromagnetically Induced Transparency (EIT) aided cooling of strontium atoms }
\author{Korak Biswas}
\affiliation{Department of Physics, Indian Institute of Science Education and Research, Pune 411008, Maharashtra, India}
\author{Kushal Patel}\affiliation{Department of Physics, Indian Institute of Science Education and Research, Pune 411008, Maharashtra, India}
\author{S. Sagar Maurya}
\affiliation{Department of Physics, Indian Institute of Science Education and Research, Pune 411008, Maharashtra, India}
\author{Pranab Dutta} 
\affiliation{Department of Physics, Indian Institute of Science Education and Research, Pune 411008, Maharashtra, India}
\author{Yeshpal Singh}
\affiliation{Midlands Ultracold Atom Research Centre, School of Physics and Astronomy, University of Birmingham, Edgbaston, Birmingham, B15 2TT, United Kingdom}
\author{Umakant D. Rapol}
\email { umakant.rapol@iiserpune.ac.in}
\affiliation{Department of Physics, Indian Institute of Science Education and Research, Pune 411008, Maharashtra, India}
%\affiliation {I-HUB Quantum Technology Foundation, Indian Institute of Science Education and Research, Pune 411008, Maharashtra, India}

\begin{abstract}

The presence of ultra-narrow inter-combination spectroscopic lines in alkaline earth elements places them as promising candidates for optical atomic clocks, quantum computation, and for probing fundmental physics. Doppler cooling of these atoms is typically achieved through two subsequent stages: the initial cooling is on the $^1S_0 \rightarrow$ $^1P_1$ transition followed by cooling using the narrow-line $^1S_0 \rightarrow$ $^3P_1$ transition. However, due to significantly lower linewidth of the second stage cooling transition, efficient transfer of atoms into the second stage becomes technically challenging. The velocity distribution of the atoms after the first stage of cooling is too broad for atoms to be captured efficiently in the second stage cooling. As a result, the capture efficiency of atoms into the second stage Magneto-Optical Trap (MOT) is low, even if the linewidth of the second stage cooling laser is artificially broadened. To address this problem, here we report the demonstration of an intermediate cooling scheme for strontium atoms ($^{88}$Sr) that exploits the phenomenon of Electromagnetically Induced Transparency (EIT) in a ‘V’ level-configuration before the atoms are transfered into the red MOT from the initial ‘blue’ MOT. Our technique significantly reduces the temperature of the atoms in the ‘blue’ MOT without a significant loss in the number of atoms, thereby increasing the loading efficiency of atoms by more than two times into the ‘red’ MOT. This technique has potential application in laser cooling of other alkaline earth atoms.

\end{abstract}

\maketitle

\section{Introduction}

Laser cooling and trapping of neutral atoms stands as a groundbreaking technique that has brought a profound transformation in atomic physics and quantum technology. This pioneering technique has paved the way for a new era of quantum metrology, quantum sensing, quantum simulation, quantum computation, and many more. This advancement has not only revolutionized the quantum phenomenon based technologies but has also given us the access to explore the exotic regime of fundamental quantum phenomena such as Bose-Einstein condensation and quantum entanglement \cite{phillipe,metcalf}. 

Laser cooling and trapping of atoms is a matter of paramount significance, particularly when dealing with atoms from the alkaline earth metal group. These atoms offer ultra-narrow optical transitions, which can be utilized for quantum information processing and extremely precise optical atomic clocks \cite{ludlow}. The conventional method of cooling alkaline earth metals involves a two-tier Doppler cooling approach to make the atoms adequately cold for various quantum experiments. They are respectively known as the initial MOT cooling phase and the narrow-line MOT cooling phase \cite{katori}. Nevertheless, due to the considerable difference between the line-widths of these MOT transitions, the inter-MOT atom transfer process becomes susceptible to a significant loss of atoms. This  is attributed to the following reason: \\
The temperature and, consequently, the velocity distribution of atoms within a MOT is directly proportional to the linewidth of the corresponding MOT transition \cite{dalibard}. Moreover, the number of atoms interacting with light within a MOT is dependent upon both the linewidth of the transition and the laser detuning. For a fixed light detuning, the lower is the linewidth, the lower is the number of interacting atoms inside a MOT \cite{ludlowthesis}. Hence, during the transfer of atoms from the initial MOT with a broader linewidth to the narrow-line MOT, a substantial majority of atoms do not interact with the laser that is precisely locked to the narrow-line transition of the secondary MOT. This causes a substantial loss of atoms from the trap during the transfer process, even if the laser detuning is artificially broadened through temporal modulation \cite{kale}.  
         
 $^{88}$Sr, an alkaline earth element, requires to undergo a sequential two-step Doppler cooling process: known as the blue MOT  $\left[(5s^2)^1S_0 \rightarrow (5s5p)^1P_1 , \text {461 nm}\right]$ and the red MOT   $\left[(5s^2)^1S_0 \rightarrow (5s5p)^3P_1 , \text {689 nm}\right]$. The linewidths of these  transitions are $2\pi \times 32$ MHz and $2\pi \times 7.5$ kHz, respectively \cite{sansonetti} which results in a poor inter-MOT transfer efficiency. In this article, we present a novel three-level cooling approach for $^{88}$Sr, aimed to reduce the temperature of the blue MOT to enhance the transfer efficiency. We have demonstrated that it is possible to reduce the temperature of the blue MOT significantly by exploiting the phenomenon called \textit{Electromagnetically Induced Transparency} (EIT) \cite{boller}. This technique effectively narrows the Doppler-broadened velocity profile of the blue MOT transition, thereby facilitating efficient loading of atoms into the narrow-line MOT.  This approach is applicable to other alkaline earth metals, ensuring a high Signal-to-Noise Ratio (SNR) throughout the experimental process.

\begin{figure*}[!ht]
	\centering
	\hspace{0.7 cm}
	\begin{subfigure}[H]{0.4\textwidth}
		\centering
		\includegraphics[width=\textwidth]{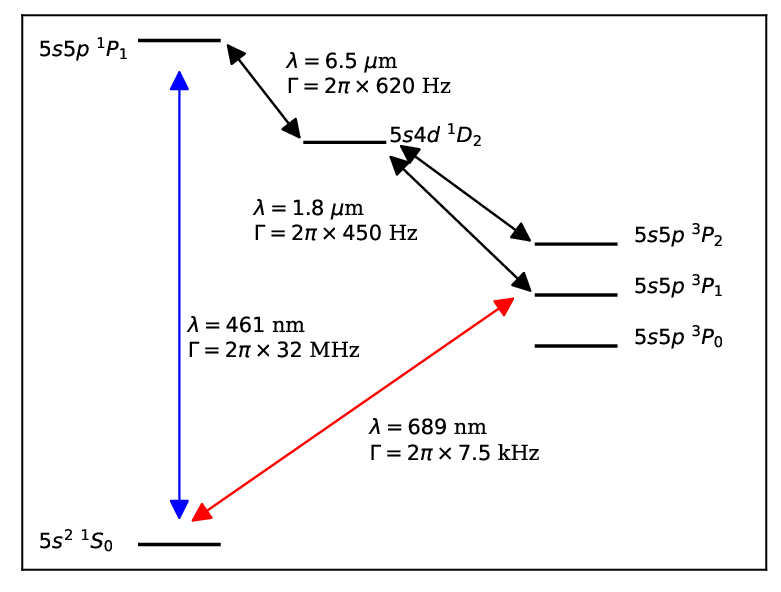}
		\caption{Low lying energy level diagram of $^{88}$Sr.}
		\label{srlevelcomp}
	\end{subfigure}
    \hspace{0.7 cm}
	\begin{subfigure}[H]{0.4\textwidth}
		\centering
		\includegraphics[width=\textwidth]{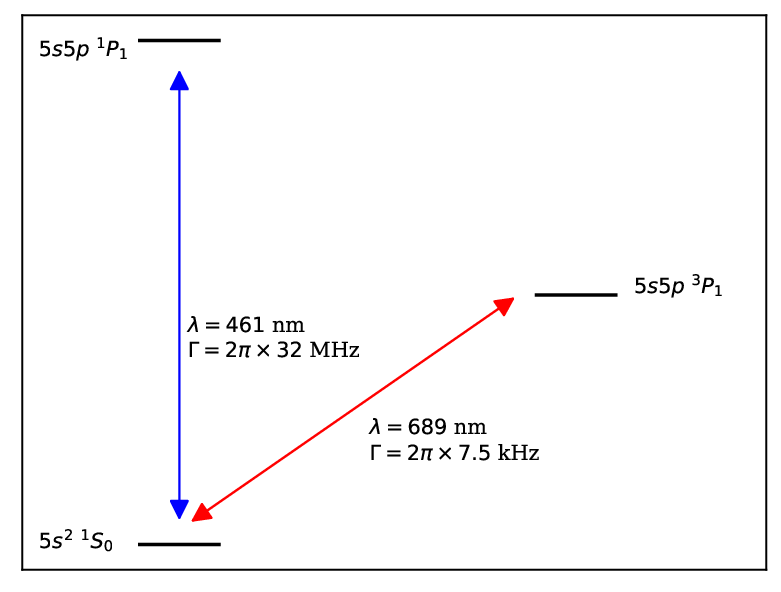}
		\caption{An approximated `V' level EIT scheme of $^{88}$Sr.}
		\label{srlevelsimp}
	\end{subfigure}
    	\begin{subfigure}[H]{0.45\textwidth}
    	\centering
    	\includegraphics[width=\textwidth]{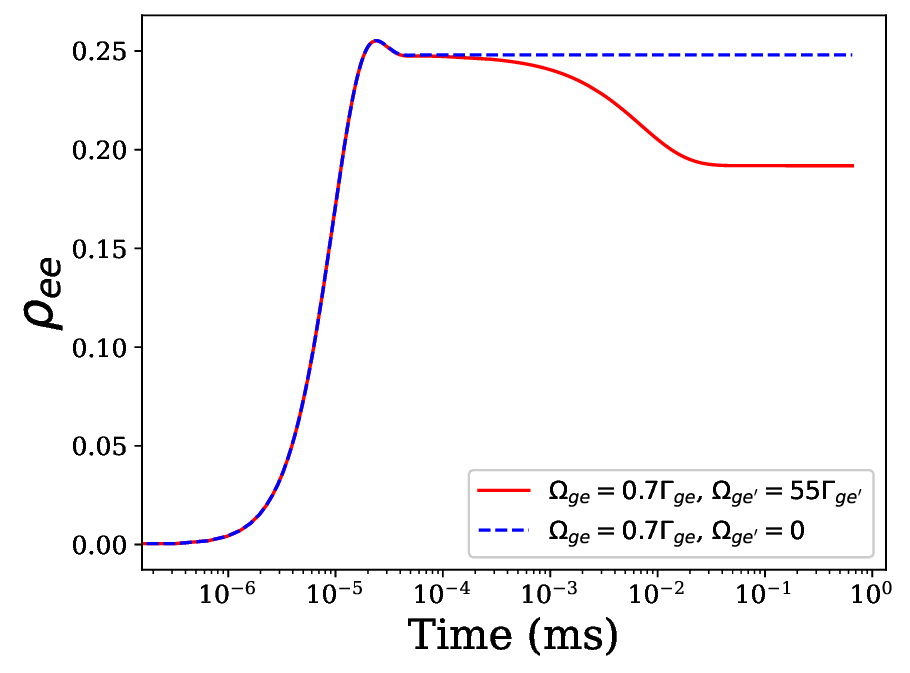}
    	\caption{Excited state population ($\rho{ee}$,$\ket{e} = \ket{^1P_1}$) of strontium atoms with zero velocity at the centre of a MOT.}
    	\label{sp}
    \end{subfigure}
\begin{subfigure}[H]{0.45\textwidth}
	\centering
	\includegraphics[width=\textwidth]{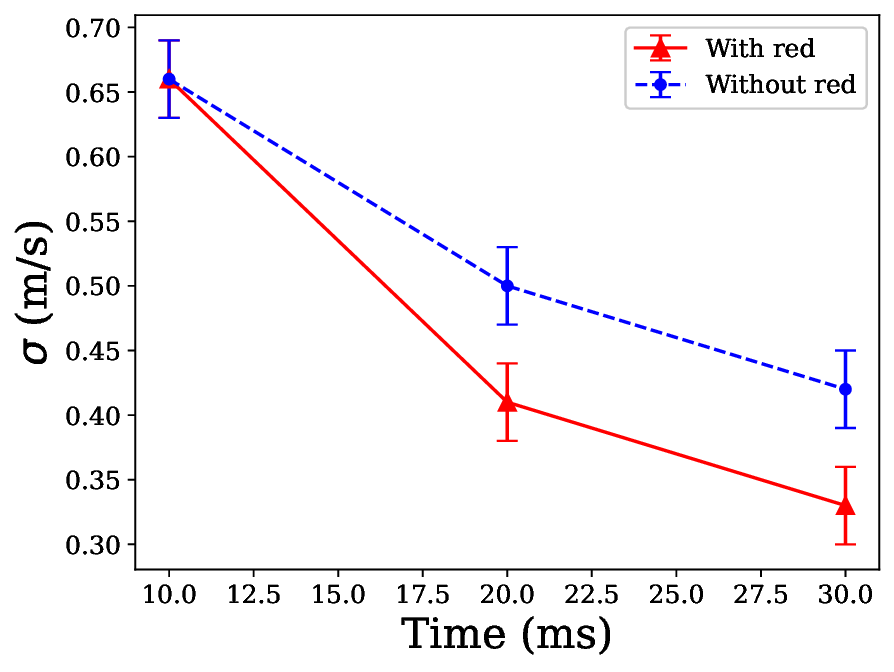}
	\caption{Standard deviation of the velocity of $^{88}$Sr atoms in a MOT in the presence (triangles) and absence (circles) of the 689 nm light.}
	\label{sigma}
\end{subfigure}
	\caption[]{ \ref{srlevelcomp} and \ref{srlevelsimp} provide the energy level diagrams for $^{88}$Sr. \ref{sp} illustrates the variation of $\rho_{ee}$ of a strontium atom, both with and without EIT. \ref{sigma} presents the standard deviation of the velocity distribution of $^{88}$Sr atoms in a MOT under two conditions: one with the intensity of blue light being ramped down in the presence of red light and the other without red light.}
	
	\label{srlevel}
\end{figure*} 

\section{Theoretical Background}
In this section, we will present a simplified theoretical model of the EIT aided cooling of the alkaline earth metals.

The 1D equation of motion of an atom in a MOT can approximately be given by the classical Langevin equation \cite{xu}.
\begin{equation} 
m\frac{dv}{dt}=-{\alpha}v-kx +\sqrt{D}\eta(t)
\label{eqn1}
\end{equation} 
 In Eq.(\ref{eqn1}) $\alpha,k=-{\partial}_{v,x}F(v,x)$, where $F$ is the linear order approximation of the actual force  $F_t= F_+(x,v)+F_-(x,v)$, $F_{\pm}(x,v)={\pm} \hbar k_{L} \Gamma \rho_{ee}$, diffusion coefficient $D=\hbar^2 k_L^2 \Gamma \rho_{ee}$ and $m,v,x,k_L,\Gamma,\rho_{ee}$ respectively denote mass of the atom, velocity of the atom, position of the atom with respect to the centre of the trap, wave-vector of the MOT laser, linewidth of the MOT transition and the fraction of the atomic population in the excited state. The function $\eta(t)$, with properties $\langle\eta(t)\rangle = 0$ and $\langle\eta(t)\eta(t+\tau)\rangle = \delta(\tau)$, characterizes the fluctuating nature of random kicks on an atom due to spontaneous emission. It is important to mention that Eq.(\ref{eqn1}) is an approximated equation and is valid only under certain limitations. 
 
 From Eq.(\ref{eqn1}), it can be shown that as $t\rightarrow \infty$, the temperature ($T$) can be given by $T = \frac{D}{k_B \alpha}$, where $k_B$ is the Boltzman constant. It is evident that by manipulating $\rho_{ee}$ one can control the temperature of the MOT. $\rho_{ee}$ can be obtained by solving the following open quantum optical Bloch equation (Eq. \ref{eqn2}) \cite{metcalf}. It is important to note here that the equation describes a non-unitary time evolution of the density operator. Consequently, it does not preserve the trace of the same. To address this, it is essential to account for the `loss of probability' at each time step while solving this equation numerically to ensure the conservation of the probability.     
 \begin{equation} 
 	\frac{\partial \hat{\rho}}{\partial t}=\frac{1}{i\hbar}[\hat{H},\hat{\rho}] - \frac{1}{2} \{\hat{\Gamma},\hat{\rho}\}
 	\label{eqn2}
 \end{equation} 

\begin{figure*}[ht]
	\centering	
	\begin{subfigure}[]{0.32\textwidth}
		\centering
		\includegraphics[width=\textwidth]{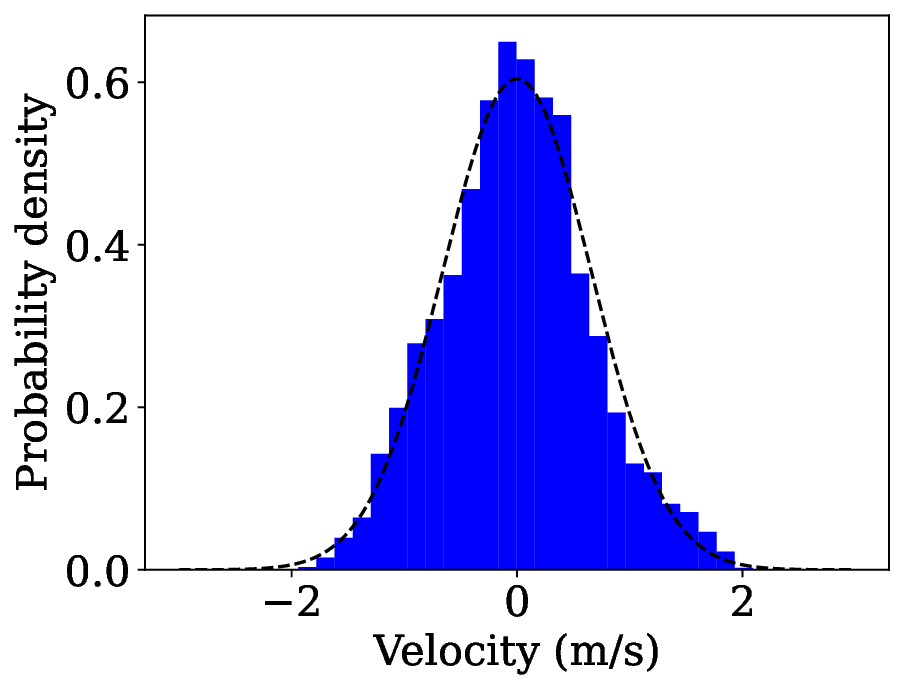}
		\caption{After 10 ms with $s_{ge} = 1.5$ and $s_{ge'} = 0$. }
		\label{H1}
	\end{subfigure}
	\centering
	\begin{subfigure}[]{0.32\textwidth}
		\includegraphics[width=\textwidth]{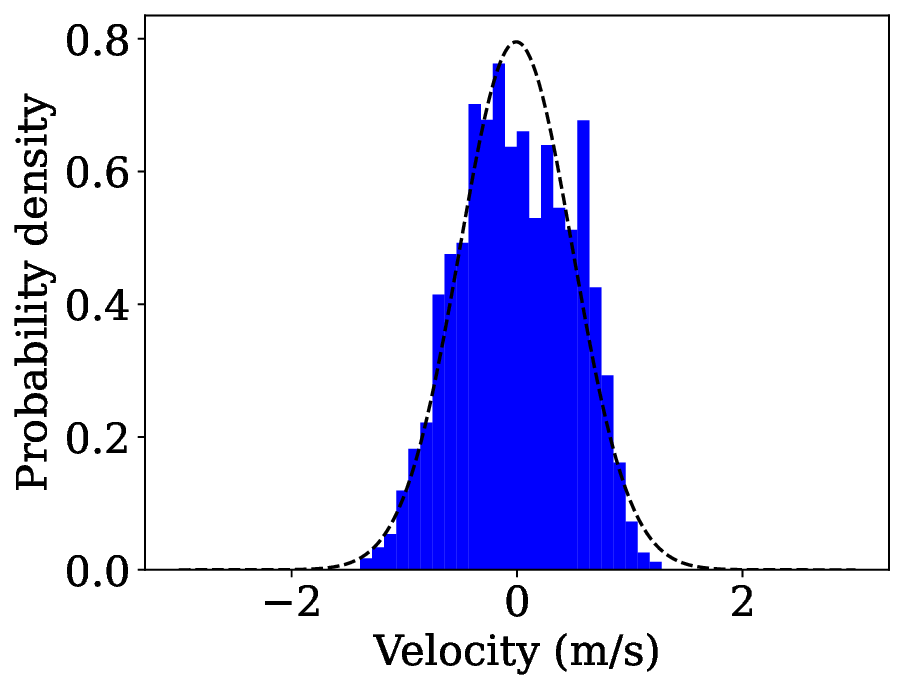}
		\caption{After 20 ms with $s_{ge} = 0.3$ and $s_{ge'} = 0$. }
		\label{H2}
	\end{subfigure}
	\begin{subfigure}[]{0.32\textwidth}
		\centering
		\includegraphics[width=\textwidth]{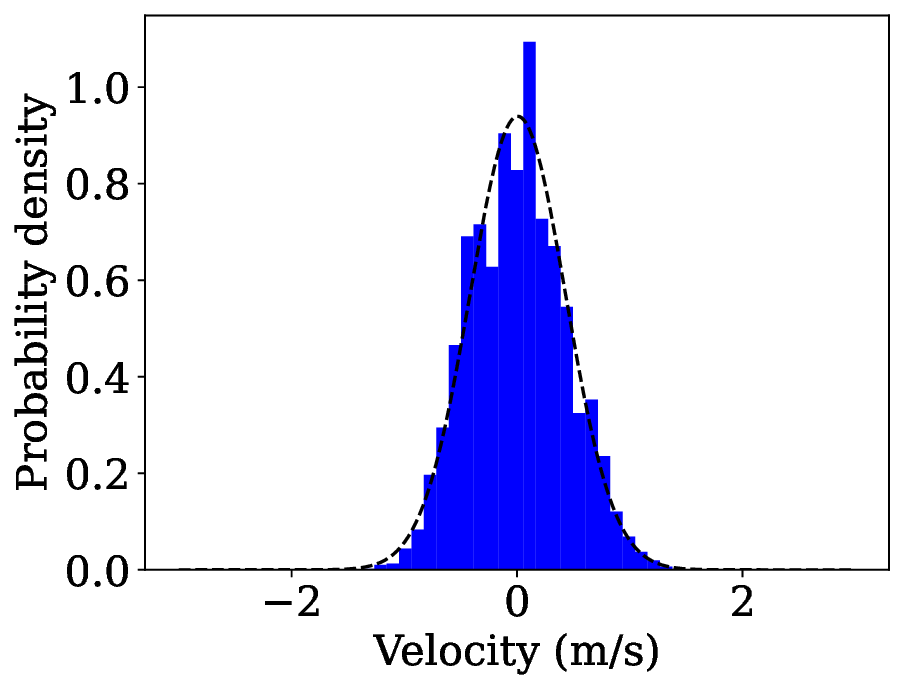}
		\caption{After 30 ms with $s_{ge} = 0.06$ and $s_{ge'} = 0$. }
		\label{H3}
	\end{subfigure}
	\begin{subfigure}[]{0.32\textwidth}
		\centering
		\includegraphics[width=\textwidth]{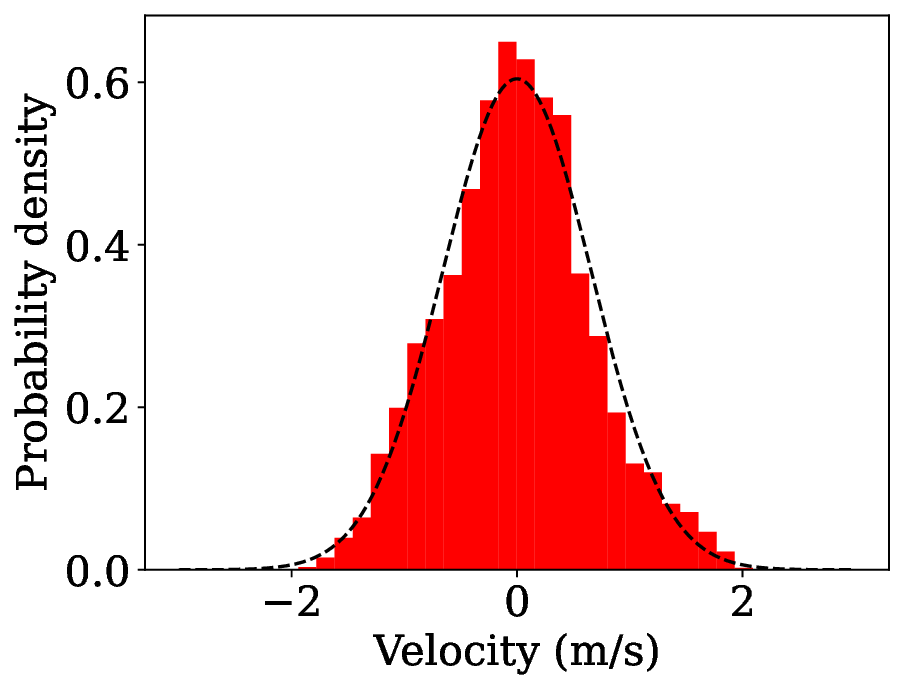}
		\caption{After 10 ms with $s_{ge} = 1.5$ and $s_{ge'} = 0$. }
		\label{H4}
	\end{subfigure}
	\centering
	\begin{subfigure}[]{0.32\textwidth}
		\includegraphics[width=\textwidth]{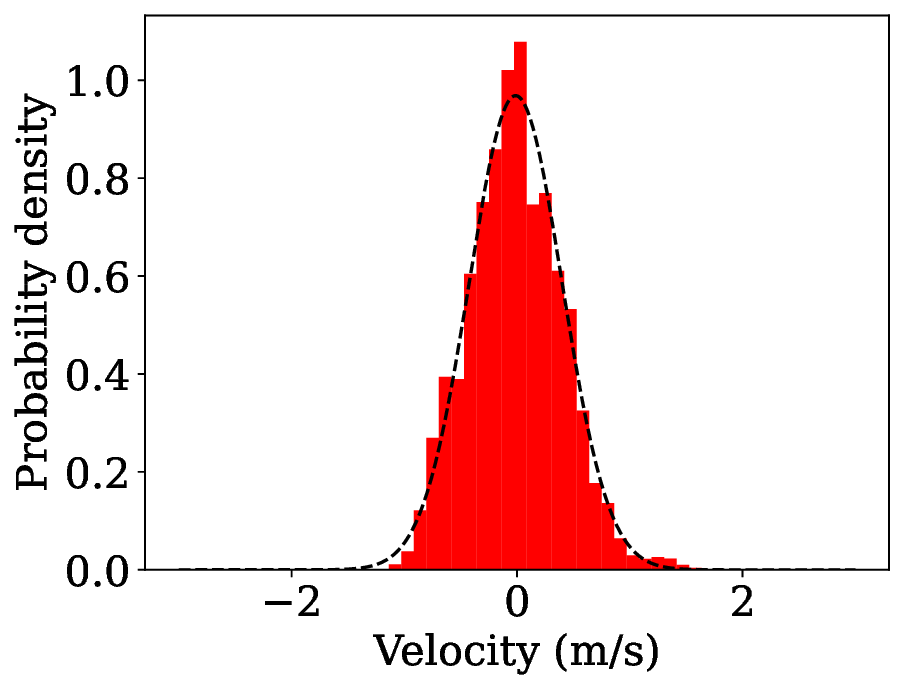}
		\caption{After 20 ms with $s_{ge} = 0.3$ and $s_{ge'} = 6000$. }
		\label{H5}
	\end{subfigure}
	\begin{subfigure}[]{0.32\textwidth}
		\centering
		\includegraphics[width=\textwidth]{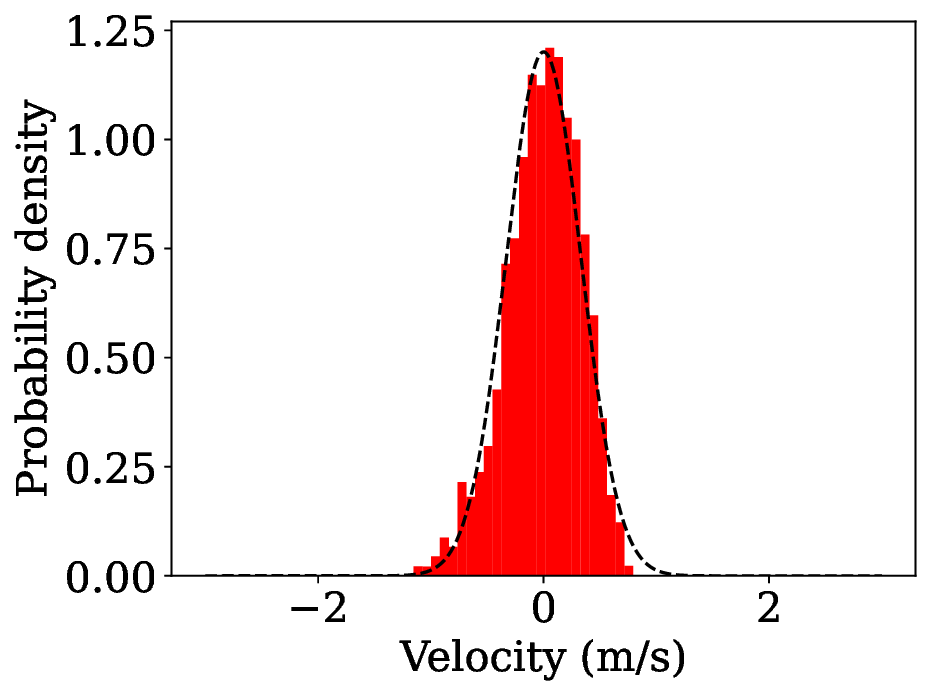}
		\caption{After 30 ms with $s_{ge} = 0.06$ and $s_{ge'} = 6000$. }
		\label{H6}
	\end{subfigure}
    \caption{Evolution of the velocity distribution of strontium atoms at different instants. This figure displays the simulated velocity distribution of $^{88}$Sr atoms within a MOT with fitted normal distribution curves. In a series of three consecutive 10-millisecond intervals, the saturation parameter of the blue light is systematically reduced from 1.5 to 0.3 and then to 0.06. Panels \ref{H1}, \ref{H2}, and \ref{H3} illustrate the velocity distribution of atoms as the blue light is progressively reduced without the introduction of red light. Where as, panels \ref{H4}, \ref{H5}, and \ref{H6} depict the velocity distribution under the same decreasing blue light conditions, but with the addition of an intense red light after the initial 10 milliseconds. Refer to FIG. \ref{sigma} for the comparison of the standard deviation of the distributions.}
    \label{Hist}
\end{figure*}

In Eq.(\ref{eqn2}) $[,]$ and $\{,\}$ represent the commutator and the anti-commutator of the variables. $\hat{\rho}$ is the density matrix and $\hat{H}$ is the Hamiltonian of light atom interaction. By implementing the rotating wave approximation (RWA) and successively performing an appropriate unitary transformation that makes $\hat{H}$ independent of time, one can write the following equations \cite{vineet} for a three level `V' system with a ground state $\ket{g}$, and a pair of excited states $\ket{e'}$ and $\ket{e}$ .

\begin{align}
	\begin{split}
		\hat{H} &= \hbar 
		\begin{pmatrix}
			0   &   {\Omega_{ge}}/{2} &   {\Omega_{ge'}}/{2}  \\
			{\Omega_{ge}^*}/{2} &   -\Delta_{ge} &  0  \\  
			{\Omega_{ge'}^*}/{2} &   0      & -\Delta_{ge'}    
		\end{pmatrix}
		\label{eqn6}  
	\end{split}    \\ 
	\begin{split} 
		\hat{\Gamma} &= \hspace{0.9cm}
		\begin{pmatrix}
			0   &   0 & 0\\
			0   &   \Gamma_{ge} & 0\\
			0   &   0    & \Gamma_{ge'} 
		\end{pmatrix}
		\label{eqn7} 
	\end{split}    \\ 
	\hat{\rho} &= \hspace{0.7cm}
	\begin{pmatrix}
		\rho_{gg}   &   \rho_{ge}  & \rho_{ge'} \\
		\rho_{eg}   &   \rho_{ee}  & \rho_{ee'} \\
		\rho_{e'g}   &   \rho_{e'e}  & \rho_{e'e'} \\
	\end{pmatrix}
	\label{eqn8}  
\end{align}

In Equation (\ref{eqn6}), $\Omega_{ge}$ and $\Omega_{ge'}$ represent the Rabi frequencies for light-assisted transitions $\ket{g} \rightarrow \ket{e}$ and $\ket{g} \rightarrow \ket{e'}$, respectively. The detunings of these transitions are denoted by $\Delta_{ge}$ and $\Delta_{ge'}$ and they include contributions from both manual detuning ($\delta = \omega_L-\omega_0$, $\omega_L$- laser frequency, $\omega_0$- resonant frequency of the atom) and Doppler detuning ($\Delta_{D}=-k_Lv$, $k_L$- wavevector of the light, $v$- velocity of the atom). When the light associated with the $\ket{g} \rightarrow \ket{e'}$ transition is switched off i.e., $\Omega_{ge'} = 0$, these equations elucidate the behavior of a conventional two-level system, where the evolution of the density matrix is solely influenced by $\Omega_{ge}$. However, by adjusting the value of $\Omega_{ge'}$, the dynamics of the system can be manipulated to reduce the value of $\rho_{ee}$ and subsequently decreasing the temperature of the atoms (FIG. \ref{sp}). This phenomenon is called EIT, as the reduction in $\rho_{ee}$ causes the atoms to exhibit partial transparency to the light associated with the $\ket{g} \rightarrow \ket{e}$ transition. 
   
To study the behavior of three-level atoms within the MOT,  it becomes crucial to explore a multi-level system as the magnetic field within the MOT splits the excited levels into a multitude of Zeeman states \cite{sakurai}.  For instance, in case of alkaline earth metals ($\ket{g} = \ket{^1S_0}$ $\ket{e} = \ket{^1P_1}$ and $\ket{e'} = \ket{^3P_1}$) , one should ideally consider a 7 level system including a pair of extra splitted levels for each of the excited states. However, in the context of a MOT, the utilization of $\sigma_+$ and $\sigma_-$ polarized light eliminates the possibilities of certain transitions, namely $\ket{^1S_0} \rightarrow \ket{^1P_1,m_f=0} $ and $\ket{^1S_0} \rightarrow \ket{^3P_1,m_f=0} $ ($m_f$ - the component of angular momentum along the quantization axis). Consequently, the system simplifies to a 5 level system. It is noteworthy that the Zeeman splitting in a MOT is inhomogeneous and is given by $\Delta_{Z}=\mu_Bm_fg\frac{dB}{dx}x$, with $\mu_B$ representing the Bohr magneton, $g$ representing the Landé g-factor of the state, and $\frac{dB}{dx}$ denoting the magnetic field gradient. This Zeeman splitting contributes to the overall detuning of the light, alongside $\delta$ and $\Delta_{D}$. With these considerations in hands, one can find out the steady-state excited level population $\rho_{ee}(t\rightarrow\infty)$ through the solution of Eq.(\ref{eqn2}) under the condition of $\frac{\partial\hat{\rho}}{\partial t}=0$ \cite{haar}, consequently calculating the temperature of the MOT. 

Alternatively, one can  also explore the dynamics of an atom in a MOT, as governed by Eq. \ref{eqn2}, over a sufficiently extended interval of time. By employing the ergodicity hypothesis, it is possible to deduce the velocity distribution of a large ensemble of atoms within the MOT. The standard deviation of this distribution serves as an indicator of the temperature of the atoms. The implementation of EIT has the notable effect of reducing the photon scattering rate ($\Gamma\rho_{ee}$) by the atoms. This reduction results in fewer random kicks per unit time on the atoms due to spontaneous emission, consequently leading to a decrease in temperature. While investigating the dynamics of an atom within a MOT, it is possible to circumvent the use of Eq. \ref{eqn1}, which represents only an approximation. Instead, one can exclusively rely on the principles of momentum conservation in atom-light interactions to evolve the dynamics. \\ \\  
 \textbf{Approximations}: The proposed EIT-assisted cooling scheme finds applicability across all alkaline atoms, owing to their `V' shape energy level configuration with $\ket{g} = \ket{^1S_0}$, $\ket{e} = \ket{^1P_1}$, and $\ket{e'} = \ket{^3P_1}$. While in practice, the scenario does not precisely emulate the `V' scheme of EIT as the transition $\ket{^1P_1} \rightarrow \ket{^3P_1}$ via $\ket{^1D_2}$ is allowed in the alkaline earth atoms. However, contributions from these transition can be neglected when $\min(\Gamma_{^1P_1\rightarrow ^1D_2},\Gamma_{^1D_2\rightarrow ^3P_1}) \ll \Gamma_{^1S_0\rightarrow ^1P_1}$ , $\Gamma_{^1S_0\rightarrow ^3P_1}$ holds true. Such an approximation remains valid for all alkaline earth metals. For instance, from the energy level diagram in FIG. \ref{srlevelcomp}, it becomes evident that this approximation remains justifiable for $^{88}$Sr atoms. As a result, one can simplify this diagram as the one depicted in FIG. \ref{srlevelsimp}. Another approximation can be taken while dealing with alkaline earth metals. When calculating the temperature of three level atoms in the MOT, one should ideally consider the diffusion coefficients arising from both transitions ($D_{^1S_0\rightarrow ^1P_1}$ and $D_{^1S_0\rightarrow ^3P_1}$). However, given that $D \propto \Gamma \rho_{ee}$ and $\Gamma_{^1S_0\rightarrow ^3P_1} \ll \Gamma_{^1S_0\rightarrow ^1P_1} $, the impact of $D_{^1S_0\rightarrow ^3P_1}$ can be neglected, unless the population in $\ket{^3P_1}$ is `sufficiently larger' than that in $\ket{^1P_1}$.

 \begin{figure*}[ht]
 	\centering
 	\begin{subfigure}[]{0.45\textwidth}
 		\centering
 		\includegraphics[height = 0.8\textwidth, width=1.03\textwidth]{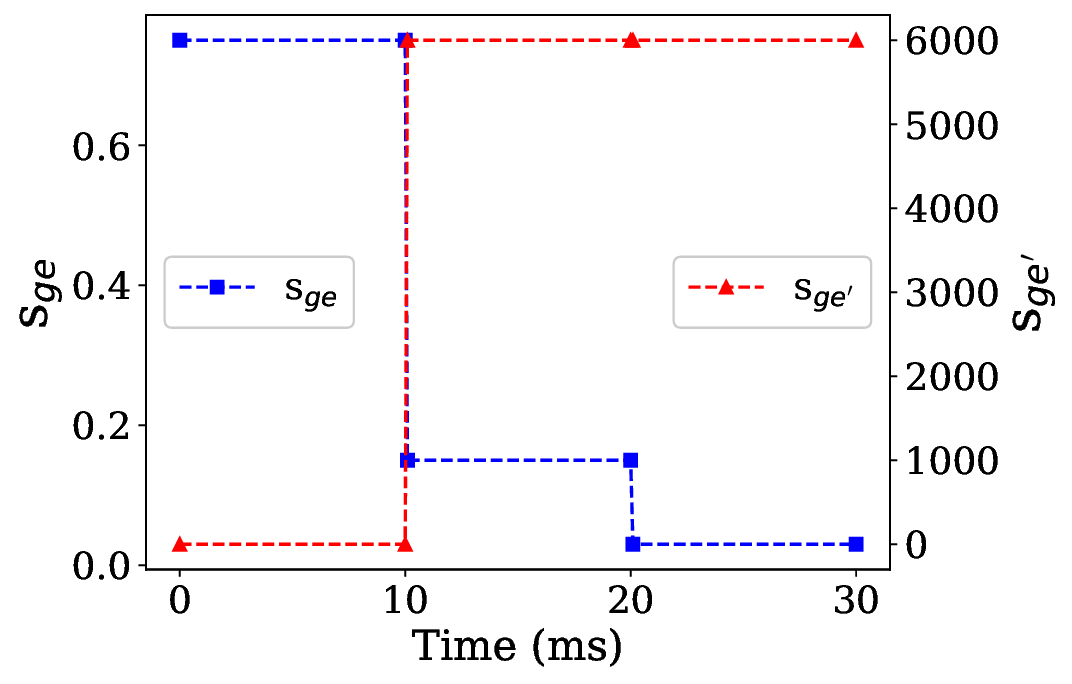}
 		\caption{Saturation parameters of the blue and the red laser during the course of the experiment.}
 		\label{T0}
 	\end{subfigure}
 	\begin{subfigure}[]{0.45\textwidth}
 		\centering
 		\includegraphics[width=\textwidth]{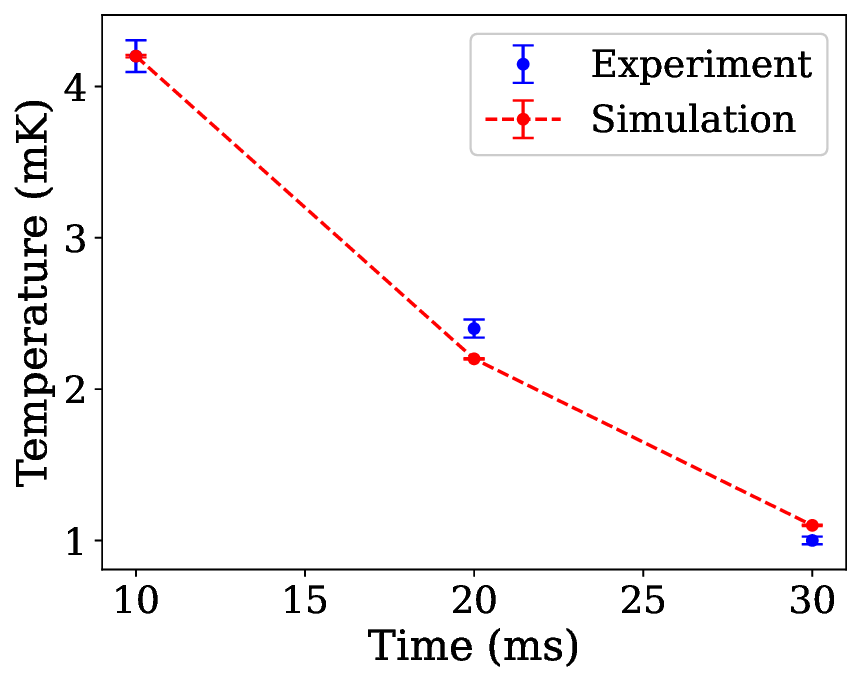}
 		\caption{Temperature vs time. Temperature decreases with time as the EIT effect intensifies.}
 		\label{T1}
 	\end{subfigure}
 	\begin{subfigure}[]{0.45\textwidth}
 		\centering
 		\includegraphics[width=\textwidth]{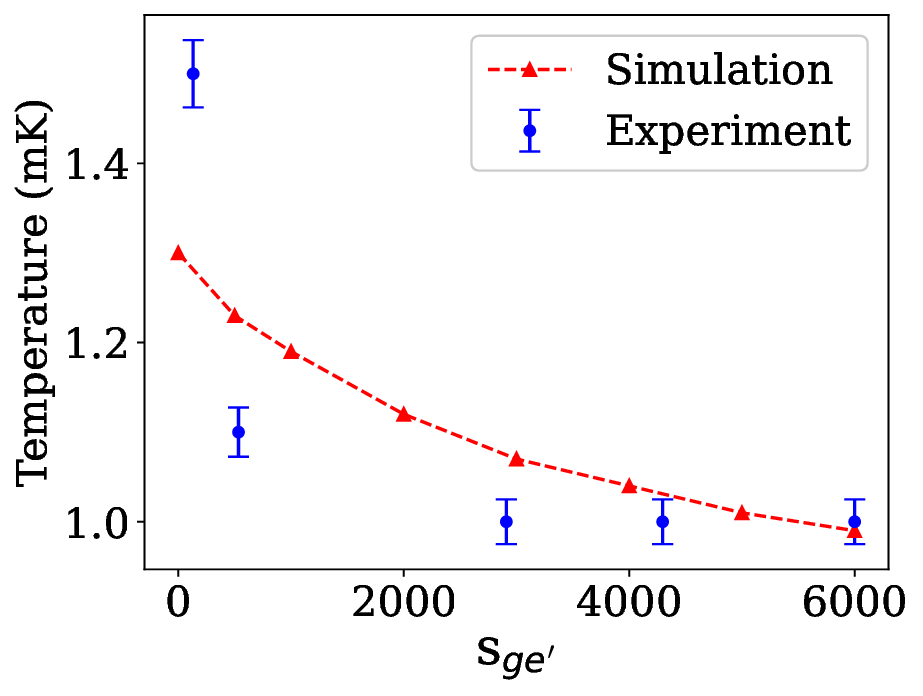}
 		\caption{Temperature vs red laser intensity after ramping down the blue intensity. ($\delta_{ge'} = -2.2$ MHz)}
 		\label{T2}
 	\end{subfigure}
 	\begin{subfigure}[]{0.45\textwidth}
 		\centering
 		\includegraphics[width=\textwidth]{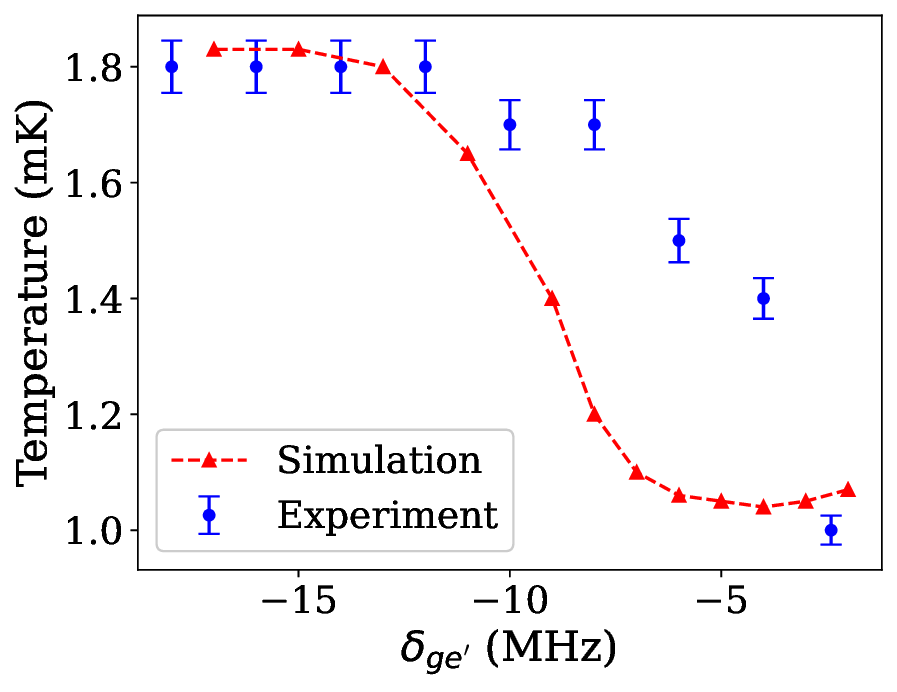}
 		\caption{Temperature vs red laser detuning after ramping down the blue intensity. (s$_{ge'} = 6000$)}
 		\label{T3}
 	\end{subfigure}
 	\begin{subfigure}[]{0.45\textwidth}
 		\centering
 		\includegraphics[width=\textwidth]{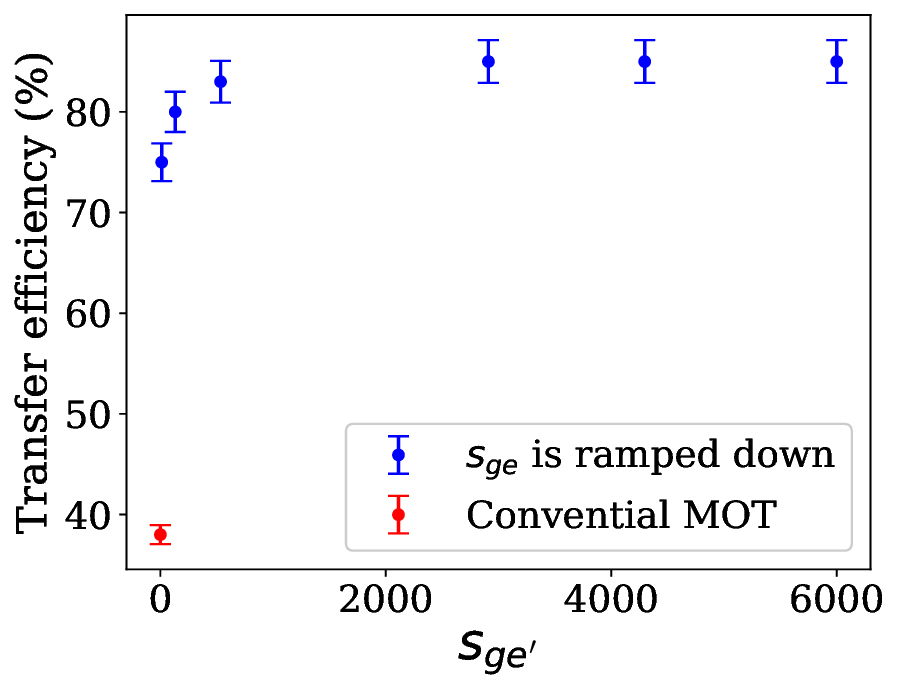}
 		\caption{Transfer efficiency vs red laser intensity after ramping down the blue intensity. ($\delta_{ge'} = -2.2$ MHz)}
 		\label{T4}
 	\end{subfigure}
 	\begin{subfigure}[]{0.45\textwidth}
 		\centering
 		\includegraphics[width=\textwidth]{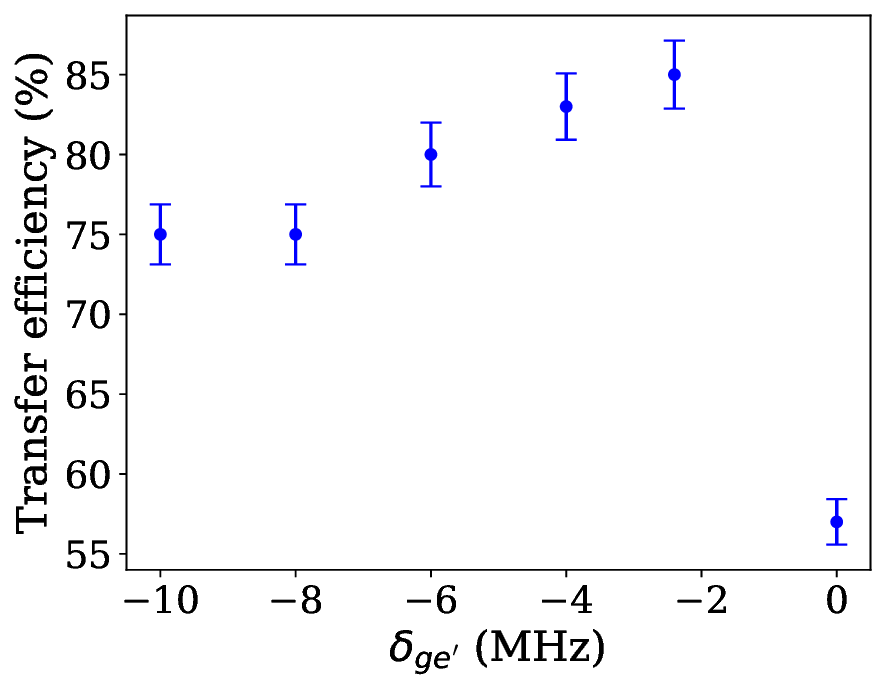}
 		\caption{Transfer efficiency vs red laser detuning after ramping down the blue intensity.  (s$_{ge'} = 6000$)}
 		\label{T5}
 	\end{subfigure}
 	%\captionsetup{justification=raggedright}
 	\caption{(a) displays the saturation parameters of the red and the blue laser at different time.  (b) shows that how MOT temperature goes down as the EIT effect increases. (c) and (d) illustrate the temperature variations of MOT with different intensities and detunings of the red laser. (e) and (f) show the transfer efficiency for different intensities and detunings of the red laser.}
 	
 	\label{Temp}
 \end{figure*}  
 
\section{Results and discussions}
From now onwards we will focus specifically on the EIT-aided cooling of $^{88}$Sr.  As previously highlighted, in case of $^{88}$Sr, a temperature reduction within the blue MOT can be achieved through the utilization of $(5s^2)^1S_0 \rightarrow (5s5p)^3P_1 $ transition. Herein, we designate the states $(5s^2)^1S_0$, $(5s5p)^1P_1 $, and $(5s5p)^3P_1 $ as $\ket{g}$, $\ket{e}$, and $\ket{e'}$ respectively. Pertaining to our system, the respective parameters are as follows: $\Gamma_{ge}= 2\pi \times 32$ MHz, $\Gamma_{ge'}= 2\pi \times 7.5$ kHz, $\lambda_{ge} = 461 $ nm (blue laser), and $\lambda_{ge'} = 689 $ nm (red laser). The Rabi frequency can be expressed as $\Omega = \sqrt{\frac{s}{2}} \Gamma $, where $s=I/I_{sat}$, with $I$ signifying the light intensity and $I_{sat}$ representing the saturation intensity of the transition. In the outlined scheme, we have $I_{sat,ge} = 42$ mW/cm$^2$ and $I_{sat,ge'} = 3$ $\mu$W/cm$^2$ \cite{yang}. Notably, the Rabi frequencies are characterized in terms of the saturation parameters $s_{ge}$ and $s_{ge'}$.

We conducted a numerical analysis of the steady-state temperature of atoms within a MOT, exploring the impact of varying intensities of 461 nm and 689 nm lasers. We observed  that the standard deviation of the velocity distribution of the atoms in the blue MOT was significantly lower than that of a conventional MOT when the intensity of the blue laser was low and the intensity of the red laser was high  (FIG. \ref{Hist}). In other words, the temperature decreased as the effect of the EIT was increased (FIG. \ref{T1}). We have also calculated the temperature of atoms in the  MOT as a function of laser detunings. We found that the temperature was minimum when the detuning of red laser was slightly lower (or higher) than 0 (FIG. \ref{T3}). 
   
We applied this technique in our experimental setup to effectively lower the temperature of the blue MOT. Following the initial conventional blue MOT cooling phase, we turned on the $689$ nm laser while gradually attenuating the power of the $461$ nm laser by controlling the RF power of an AOM (Acousto-Optic Modulator) configured in a double-pass arrangement.We systematically lowered the power of the blue laser over a sequence of three discrete time intervals, each spanning 10 ms (FIG. \ref{T0}). The saturation parameter values corresponding to the blue laser for these intervals were set at 1.5, 0.3, and 0.06, respectively. Simultaneously, the saturation parameter values for the red laser were set at 0, 6000, and 6000, respectively. Using the \textit{Time of Flight} (ToF) method, we measured the rate of expansion of the atomic cloud and inferred that the temperature of the blue MOT was reduced substantially due to the effect of EIT (FIG. \ref{T1}). It is worth mentioning here that a gradual reduction in the power of the blue laser is necessary as an abrupt decrement in laser power causes a sudden reduction in the MOT force leading to the loss of significant number of atoms.

Furthermore, we investigated the influence of the red laser's intensity while maintaining a constant blue laser intensity. The reduction in the red laser's intensity led to an attenuation of the EIT effect, subsequently resulting an increment in the MOT's temperature (see FIG. \ref{T2} for illustration).
Similarly, we explored the detuning effect of the red laser (refer to FIG. \ref{T3}) on the temperature. We found that the temperature reached its minimum point at a detuning slightly below zero, at which the maximum fraction of the atoms were in resonance with the red light. It is worth mentioning that all the data found from the experiments are in good agreement with the theoretical predictions of the approximated model we described in one of the earlier scetions. 

Implementing the aforementioned technique, we have achieved a significant decrease of temperature of atoms in the MOT --  more than $75\%$ (FIG. \ref{T1}). Since a lower temperature of the atomic cloud leads to the compression of the atomic velocity distribution, a greater number of atoms in the blue MOT interacted with the red laser leading to a higher inter MOT transfer efficiency (see FIG. \ref{T4} and \ref{T5} for illustration). Our observations indicated a significantly enhanced transfer efficiency compared to the conventional approach. Employing the EIT method, we achieved a transfer efficiency exceeding $85\%$, a notable improvement compared to $39\%$ attained without utilizing the EIT cooling.

\section{Conclusion}
 We demonstrated an innovative three-level cooling approach for strontium atoms, harnessing the advantageous effect of EIT. This technique significantly reduced the temperature of the blue MOT, leading to a notable enhancement in the efficiency of inter-MOT atom transfer. This method can be approximately modeled with the 'V' scheme of EIT, and our experimental results are in good agreement with the values calculated through theoretical predictions of the approximate model. Notably, this cooling method can be applied across the all the elemnts belonging to the alkaline erath metal group as they share similar level structures. By applying this approach, we can effectively amplify the inter-MOT loading efficiency for these elements, consequently elevating the SNR for various quantum experiments.        
\vspace{0.1cm}
\section{Acknowledgments}
The authors would like to thank the Department of Science and Technology, Govt. of India, for funding through the Quantum Enabled Science and Technology (QuEST) program and the I-HUB Quantum Technology Foundation through the National Mission on Interdisciplinary Cyber-Physical Systems (NM-ICPS). SSM and PD acknowledge research fellowship from Council of Scientific \& Industrial Research (CSIR) Govt. of India. We also acknowledge the support of the National Supercomputing Mission (NSM) for providing computing resources of ‘PARAM Brahma’ at IISER Pune, which is implemented by C-DAC and supported by the Ministry of Electronics and Information Technology (MeitY) and Department of Science and Technology (DST), Govt. of India.

%\bibliographystyle{ieeetr} % We choose the "plain" reference style
%\bibliography{ref} % Entries are in the refs.bib file

\end{document}